\documentclass[reprint,amsmath,amssymb,aps,superscriptaddress,prd,nofootinbib, floatfix]{revtex4-2}
\usepackage{geometry}
\usepackage{graphicx}
\usepackage{mathtools}
\usepackage{siunitx}
\usepackage{xcolor}
\usepackage[normalem]{ulem}
\usepackage{url}
\usepackage{hyperref}
\usepackage{soul}
\usepackage{booktabs}
\usepackage{multirow}

\begin{document}

\title{Addressing the challenges of detecting time-overlapping compact binary coalescences}

\author{Philip Relton}
\email[Correspondence email address: ]{reltonpj@cardiff.ac.uk}
\affiliation{Gravity Exploration Institute, Cardiff University, Cardiff CF24 3AA, United Kingdom} % Phil & Vivien
\author{Andrea Virtuoso}
\affiliation{Università di Trieste, Dipartimento di Fisica, I-34127 Trieste, Italy; INFN Sezione di Trieste, I-34127 Trieste, Italy} % Andrea (V)
\author{Sophie Bini}
\affiliation{Università di Trento, Dipartimento di Fisica, I-38123 Povo, Trento, Italy; INFN, TIFPA, I-38123 Povo, Trento, Italy} % Andrea (M) & Sophie
\author{Vivien Raymond}
\affiliation{Gravity Exploration Institute, Cardiff University, Cardiff CF24 3AA, United Kingdom} % Phil & Vivien
\author{Ian Harry}
\affiliation{University of Portsmouth, Portsmouth, PO1 3FX, United Kingdom} % Ian
\author{Marco Drago}
\affiliation{Università di Roma La Sapienza, I-00185 Roma, Italy; INFN, Sezione di Roma, I-00185 Roma, Italy} % Marco
\author{Claudia Lazzaro}
\affiliation{10 Università di Padova, Dipartimento di Fisica e Astronomia, I-35131 Padova, Italy; INFN, Sezione di Padova, I-35131 Padova, Italy} % Claudia
\author{Andrea Miani}
\affiliation{Università di Trento, Dipartimento di Fisica, I-38123 Povo, Trento, Italy; INFN, TIFPA, I-38123 Povo, Trento, Italy} % Andrea (M) & Sophie
\author{Shubhanshu Tiwari}
\affiliation{Physik-Institut, University of Zurich, Winterthurerstrasse 190, 8057 Zurich, Switzerland} % Shubhansu

\date{\today}

%----------------------------------------------------------
%----------------------- Abstract -------------------------
%----------------------------------------------------------
\begin{abstract}
  Standard detection and analysis techniques for transient gravitational waves make the assumption that detector data contains, at most, one signal at any time. As detectors improve in sensitivity, this assumption will no longer be valid. In this paper we examine how current search techniques for transient gravitational waves will behave under the presence of more than one signal. We perform searches on simulated data sets containing time-overlapping compact binary coalescences. This includes a modelled, matched filter search (\texttt{PyCBC}), and an unmodelled coherent search, coherent WaveBurst (\texttt{cWB}). Both of these searches are used by the LIGO-Virgo-KAGRA collaboration \cite{abbott2021gwtc}. We find that both searches are capable of identifying both signals correctly when the signals are dissimilar in merger time, $|\Delta t_c| \geq \SI{1}{s}$, with \texttt{PyCBC} missing only $1\%$ of the overlapping binary black hole mergers it was provided. Both pipelines can find signal pairings within the region $|\Delta t_c| < \SI{1}{s}$. However, clustering routines in the pipelines will cause only one of the two signals to be recovered, as such the efficiency is reduced. Within this region, we find that \texttt{cWB} can identify both signals. We also find that matched filter searches can be modified to provide estimates of the correct parameters for each signal.
\end{abstract}

\maketitle

%----------------------------------------------------------
%--------------------- Introduction -----------------------
%----------------------------------------------------------
\section{Introduction} \label{intro}
In 2015 the Advanced Laser Interferometer Gravitational-Wave Observatory (aLIGO) interferometers observed the merger of two black holes \cite{abbott2016gw150914}. Since then, there have been over 90 transient gravitational waves observed ~\cite{abbott2021gwtc, nitz2021}. The rate of these mergers is given by the rate of gravitational wave from astrophysical sources and the observing range of the detector. As the current network of detectors are improved ~\cite{aso2013interferometer, LIGOScientific2014pky, acernese2014advanced, kagra2019kagra, abbott2020prospects}, and future detectors such as The Einstein Telescope (ET) \cite{punturo2009einstein} and Cosmic Explorer \cite{reitze2019cosmic} begin observations, this observing range will significantly increase. This will lead to multiple transient signals being present in the detector concurrently, most likely before the end of the decade \cite{relton2021parameter}, with most transient signals overlapping another in the next generation of observatories.

Previous investigations into the case of time-overlapping compact binary coalescences (CBCs), have focused on how parameter estimation techniques will perform. Their findings indicate that the parameters of modelled signals should be recoverable, unless the two signals overlap strongly with a small relative merger time ~\cite{relton2021parameter, samajdar2021biases, himemoto2021impacts, antonelli2021noisy, pizzati2022toward}. This is stronger if the signals have similar signal-to-noise ratios (SNRs) \cite{relton2021parameter}, or chirp masses, mostly due to similar frequencies in the signals \cite{himemoto2021impacts}. In the case where overlapping astrophysical signals are misidentified as a single signal, the data will be passed on to these parameter estimation methods. As these methods aim to minimise their residuals, it is possible that the overlap would not be noticed. Any resulting oddities would likely be attributed to other causes such as glitches, precession or eccentricity ~\cite{abbott2017gw170817, davis2022subtracting, payne2022curious}.

Current methods of gravitational wave detection rely upon the assumption that only a single signal is present in the detector and, in most algorithms, that this signal can be matched to modelled signal waveforms. As this assumption will break for time-overlapping transients, we have investigated how current search pipelines will behave in their presence.

Our investigation was performed by injecting pairs of overlapping CBCs into the coloured-Gaussian noise of a three detector, LIGO-Virgo, network ~\cite{abbott2008advligo, acernese2014advanced}. We then performed an offline search of these data sets using a modelled search, \texttt{PyCBC}) ~\cite{allen2005chi, allen2012findchirp, dal2014implementing, usman2016pycbc, nitz2017detecting, alex_nitz_2020_4309869}, and an unmodelled search, \texttt{cWB} ~\cite{klimenko2016method, klimenko_sergey_2021_5798976}. We investigated how the parameters of the signal pairings affected whether the signals were detected and how accurate the recovered parameters of those signals were.

Section \ref{method} outlines how the data were generated, and how each search works. Section \ref{results} covers the findings of how signal pairings were detected and the effect of the overlap on recovered parameters. Section \ref{Identification} shows how these search methods could lead to successful overlap identification and separation. Section \ref{discussion} contains a summary of our findings, followed by an appendix.

%----------------------------------------------------------
%------------------------ Method --------------------------
%----------------------------------------------------------
\section{Method} \label{method}
\subsection{Overview} \label{methodoverview}
For this study we consider two types of CBCs, binary black hole mergers (BBH) and binary neutron star mergers (BNS). Our main interest in including these two forms of signal was to see how current search methods behave under the presence of overlapping CBCs of different visible durations and frequency profiles. We did not include neutron star-black hole signals, which should have similar observable durations \cite{abbott2021observation} to low mass BBH signals, as these events are rare and less relevant to this study. Their presence in overlaps should be considered in future studies. We also did not include tidal deformability in the BNS signals, as it should have little effect on the length and frequency of the signal \cite{dietrich2019matter}. Three separate investigations were performed for this study, each examining one of three combinations of CBC overlap: BBH+BBH, BNS+BNS and BNS+BBH.

\subsection{Data generation} \label{datagen}
Each investigation in this study included generating ten days of coloured-Gaussian noise for a three detector network. This length of time was chosen to give a reasonable estimate of false alarm rate of significant signals and to allow a reasonable number of events to be studied. The detectors were added as two advanced LIGO detectors at design specification in the locations of LIGO-Hanford and LIGO-Livingston ~\cite{abbott2008advligo, aligo2018psd} and a third detector with an Advanced Virgo sensitivity in the position of Virgo \cite{acernese2014advanced}.

For each of the three studies mentioned in Section \ref{methodoverview} we made three injection sets. The first two only included individual, non-overlapping signals from the pairings, labelled as $\mathrm{SINGLES}_{\mathrm{A}}$ and $\mathrm{SINGLES}_{\mathrm{B}}$. This provides comparison data to see whether each signal is detectable alone. The final injection set, labelled as PAIRS, contained both Signal A and Signal B as time-overlapping pairs.

Signals were injected into the data internally in each search pipeline. BBH signals were generated using the SEOBNRv4PHM approximant \cite{ossokine2020multipolar} to allow for the inclusion of spin precession and higher modes. BNS signals were generated with the SEOBNRv4 \cite{bohe2017improved} approximant; this only allows for aligned spin and, does not include neutron star tidal deformability.

With the exception of merger time, the extrinsic parameters of both Signal A and B were drawn from uniform distributions in sky locations, phase, polarisation and binary inclination. Each signal was generated from $\SI{9}{Hz}$, just below the noise cutoff of $\SI{10}{Hz}$, to avoid discontinuities in the data. Further details of intrinsic parameters, merger times and luminosity distances are given in the subsections below.

This distribution is designed to be as close to a true astrophysical distribution, but to also allow reasonable study of possible overlapping CBC cases. As such our study is aimed at the methodology of the pipelines and not at providing an astrophysical study of such events.

\subsubsection{Masses and spins} \label{intrinsicselection}
The masses of the two objects in a binary were drawn in pairs. BBH mass pairing were drawn from the most recent binary mass distribution estimation by the LIGO-Virgo collaboration. The chosen mass distribution was the PowerLaw+Peak distribution covering primary masses in the range $[5,100]\, M_{\odot}$ and mass ratios in the range $[0.1,1]$ \cite{abbott2021population}. As the detectors have some bias as to the events they will detect, we included a biasing factor on the mass distribution. Using the power spectral density (PSD) of the advanced LIGO detectors \cite{aligo2018psd}, and the \texttt{inspiral-range} \texttt{PYTHON} package ~\cite{chen2021distance, gwinc-inspiralrange}, we estimated viewing ranges for each binary pairing in the mass distribution grid. These distance estimates were used to generate volume estimates and were then multiplied by the per volume merger rate to estimate a true merger rate of each signal. This was then normalised to produce a weighted distribution to bias the astrophysical merger rates, by mass, to the observable merger rate. This produces binaries with more even mass ratios and higher masses, as is expected in these detectors.

The neutron star (NS) binary mass range is still an area of active research \cite{abbott2021population, landry2021mass}, with observed and predicted maximum NS masses differing ~\cite{Rhoades1974fn, rezzolla2018using}. As such, for BNS mergers we defined a mass range of $[1.14,3]\, M_{\odot}$ and drew uniformly across this. This is broadly consistent with both gravitational wave and electromagnetic observations. No mass distribution biasing was performed due to the smaller luminosity distances drawn for these binaries.

BBH spins were drawn from the spin distributions estimated by the PowerLaw+Peak mass model from the most recent estimates by the LIGO-Virgo collaboration \cite{abbott2021population}. This includes both aligned and in-plane spins with azimuthal orientations between $[\frac{-\pi}{2}, \frac{\pi}{2}]$, polar angles between $[0,2\pi]$ and magnitudes ranging between $[0,0.99]$ in dimensionless magnitude. Parameter estimation on time-overlapping signals has shown that highly overlapping pairs can mimic precession effects in the waveform \cite{relton2021parameter}, so it is interesting to consider how well searches will observe such events. Spins were drawn independently of binary masses. BNS signals were given aligned only spins with magnitude range $[-0.05,0.05]$, based on observations of BNS systems ~\cite{lynch2012timing, abbott2017gw170817, abbott2020gw190425}.

\subsubsection{Distances and times} \label{extrinsicselection}
The distance of events was drawn from a second order power law between $[200, 1300]\, \SI{}{Mpc}$ for BBH systems and $[5, 200]\, \SI{}{Mpc}$ for BNS systems. These values were set to ensure a reasonable number of visible events in the injection sets, without too many events being too quiet to observe, $\mathrm{SNR} < 5$, or overwhelmingly loud, $\mathrm{SNR} > 50$. These ranges are broadly consistent with observations in current detectors \cite{abbott2021gwtc}. While redshift does have an effect at these distances it will not change any conclusions drawn from this study, and therefore was not included in these injections. For consistency, all detector frame masses are identical to the source frame masses.

The merger times of Signal A events were drawn uniformly across the ten days of simulated data. To avoid overlaps of $N_{signals}>2$, these times were spaced out within this data set such that no mergers were within $60$ seconds of the start of when the next pairing reached $\SI{10}{Hz}$. The merger times of Signal B events were then selected by generating a waveform for each Signal A, calculating the time of first visibility at $\SI{10}{Hz}$ and that of the merger, and drawing a time uniformly between these values. To ensure observations of close relative merger times, some pairings in the BNS+BNS and BNS+BBH runs the times were drawn within two seconds of the merger. The BNS studies with narrower merger time separations were performed as a separate ten day segment, which does not affect the estimated false alarm rates in those runs.

As lower mass signals, such as BNS inspirals, are in-band in the detector for much longer than lower mass, BBH signals, this leads to many fewer signals in the BNS study. However, we did not increase the length of the BNS studies beyond ten days as this would have drastically increased the computational expense.

\subsection{Searches} \label{mastersearch}
\subsubsection{\texttt{PyCBC}) search} \label{searchpycbc}
\texttt{PyCBC}) is a matched-filter search pipeline for the detection of CBC signals in a wide parameter space ~\cite{allen2005chi, allen2012findchirp, dal2014implementing, usman2016pycbc, nitz2017detecting, alex_nitz_2020_4309869, nitz2021}. The triggers are generated in coincidence with the network of detectors by correlating the data with templates. The \textit{bank} of templates used in this study is exactly the same as the one used for \texttt{PyCBC-broad} in GWTC-3 \cite{abbott2021gwtc}. In this work we have employed a slightly modified version of \texttt{PyCBC-broad} which was used for the GWTC-3 catalog \cite{abbott2021gwtc}. These modifications were made in order to accommodate the complexity induced in the signal space due to time overlap. In particular, for this work the ``clustering window" of the \texttt{PyCBC}) search is modified. This is discussed in more detail in Section \ref{findingcriteria}. 

Background estimates are generated by time shifting the data of the three detectors by more than the time of flight between the detectors and a background trigger list is obtained. This process is repeated until enough events are acquired to measure a false alarm rate (FAR) of 1 per year ($\SI{1}{yr^{-1}}$). The injections are then processed and ranked according to the background. This process is same for both the pipelines used in this study, however the detection statistics between the two pipelines are very different. 
 
\texttt{PyCBC}) uses detection statistics based on the matched-filter SNR, with information about the event rate and the background rate incorporated as described in ~\cite{nitz2021, Davies:2020tsx}. 

\subsubsection{\texttt{cWB} search} \label{searchcwb}

\texttt{cWB} is an unmodelled analysis pipeline which detects and reconstructs gravitational wave (GW) signals without assuming any waveform model ~\cite{klimenko2016method, klimenko_sergey_2021_5798976, drago2021coherent}. \texttt{cWB} decomposes each interferometer data into a time-frequency (TF) representation using Wilson-Daubechies-Meyer wavelets \cite{necula2012transient}. 
Each wavelet amplitude is normalised by the corresponding detector amplitude spectral density, \texttt{cWB} then selects those wavelets having energy above a fixed threshold. Finally, clusters from different detectors are combined coherently into a likelihood function, which is maximised with respect to the sky location.

From the likelihood we define the statistical quantities to distinguish between a possible GW signal and glitches coming from the noise. The first being the coherent energy $E_{c}$, which represents the coherent contribution of the likelihood by cross-correlating data from different detectors.
Another key statistic is the null, or residual, noise energy $E_{n}$, which is given by subtracting the likelihood from the whitened data energy.

From coherent and null energy we can define two further quantities. The first is the penalty, $\chi^{2}$, defined as the null energy divided by the number of independent wavelet amplitudes used for describing the detected event. The second is the correlation coefficient $c_c$:
\begin{equation}
    c_{c} \,=\, \frac{\left|E_{c}\right|}{\left|E_{c}\right|+E_{n}}
\end{equation}

This estimates the coherence of the data among different detectors: when $c_c \simeq 1$ ($\left|E_{c}\right|\gg E_{n}$) there is a high coherence and it is likely that there is an astrophysical signal, while when $c_c \simeq 0$ ($E_{n}\gg\left|E_{c}\right|$) the data is incoherent and so is less likely to contain an astrophysical signal.

The correlation coefficient and the penalty are used in post-production analysis for recognising non-Gaussian noise transients, commonly referred to as ``glitches", which could trigger the pipeline, even if they are not GW events. In this work we applied the same cuts used for O3 analysis, these are $c_c>0.7$ and $\log_{10}(\chi^{2})<0.2$ \cite{abbott2021gwtc}.

\subsection{Finding signals} \label{findingcriteria}
Once the searches were performed, trigger sets were recovered for each of the three injection sets in each run. Injections were marked as found if there was a trigger found in the trigger sets within a defined time separation at merger. These separations differed between the two pipelines.

As a matched filter search, \texttt{PyCBC}) returns a time for each trigger that corresponds to the visible end of the signal in the data. This was directly matched to the injected signals merger time. If the injected signal had a \texttt{PyCBC}) trigger with an end time in the range $\pm0.1$ seconds then the injection was counted as found by the pipeline.

A slight problem arises for signal pairings in which the mergers are within $\pm0.1$ seconds. \texttt{PyCBC}) often finds that several templates match the same signal, particularly for significant triggers. To avoid this a clustering window is specified for the run. This window, set to $\pm1$ second for our study, will reject all but the most significant trigger within the window. If signals overlap in this region, then only one trigger will be returned for the pairing. This situation is covered in more detail in Section \ref{cbcresults}.

\texttt{cWB}'s approach of searching for regions of excess power requires a different approach. The standard returned time is the mean time, weighted with energy, but for this study it is more suitable to use the end time of the reconstructed waveform as estimated by \texttt{cWB}. This allows for the reconstructed waveforms to be counted correctly. Initial applications of the \texttt{PyCBC}) time constraints led to a large number of triggered injections being rejected. A window for signal finding of injected/reconstructed signal end time was set to $\pm2.5$ seconds. This time allows for the best recovery rate of \texttt{cWB} triggers, without missing separate signal pairings.

%----------------------------------------------------------
%------------------------ Results -------------------------
%----------------------------------------------------------
\section{Results} \label{results}
\subsection{Bias regions for overlapping signals} \label{OLregions}
Previous studies of time-overlapping CBCs have focused on their effect on the parameter estimation of the underlying signals ~\cite{relton2021parameter, samajdar2021biases, himemoto2021impacts, antonelli2021noisy, pizzati2022toward}. By considering the findings of these studies, we define three regions in which the presence of a second signal affects the recovered parameters of the primary signal:

\textbf{Strong bias}: This is the region in which the two signals most strongly affect each other, recovered parameters will be significantly biased away from their true values. Largely this is bound by the separation of merger times between the signals. This boundary is not consistent between all studies, to be conservative we define this as a merger time separation of $|\Delta{t_c}| \leq \SI{0.5}{s}$ for BBH+BBH overlaps. However, some studies ~\cite{himemoto2021impacts, pizzati2022toward}, indicate that the similar region for BNS+BNS overlaps is closer at $|\Delta{t_c}| \leq \SI{0.01}{s}$. The literature has not defined any such value for BNS+BBH overlaps, for this study we have applied boundaries to be the same as BNS+BNS overlaps. This region is smaller, likely due to the number of clean cycles in the later merging BNS, due to their high merger frequency, unlike BBH mergers which merge at lower frequency.

\textbf{Weak bias}: In this region signals are often recovered with slightly biased, but broadly correct, parameters. Considering the literature we define this region as $0.5 < |\Delta{t_c}| \leq \SI{2}{s}$, with the lower limit varying for overlaps containing a BNS merger.

\textbf{Negligible bias}: In this region the signals are dissimilar enough to not cause any noticeable bias in the recovered parameters. Pairings in this region should both be recovered. The bias in one signal, caused by the presence of the other, should be small enough to not negatively impact the results. We define this region to be for merger time separations of $|\Delta{t_c}| > \SI{2}{s}$.

Parameter estimation studies also indicate that the bias will be negligible if the ratio of the SNRs of the signals is particularly uneven, one signal being at least greater than three times louder than the other \cite{relton2021parameter}. Pairings in this category are likely to fall into the negligible bias region, regardless of merger time proximity. Some indication of relative chirp mass, and therefore waveform frequency range, is likely to also have an effect \cite{himemoto2021impacts}. If signals differ significantly in chirp mass, $\mathcal{M}$, then the bias may be smaller as the signals are dissimilar in frequency at merger. While we include signal overlaps of differing chirp mass in the study, we do not examine how this effect is present in trigger selection.

\begin{table*}[ht]
    \centering
    \begin{tabular}{cllll}
        \toprule\toprule
        {} & \multicolumn{3}{c}{$N_{Overlaps}$ by region} \\
        {Overlap configuration} & Strong & Weak & Negligible \\
        \midrule \vspace{1mm}
        BBH+BBH & $5.6^{+9.1}_{-3.3}$ & $17.0^{+27.0}_{-9.8}$ & $13000.0^{+8300.0}_{-4900.0}$ \\ \vspace{1mm}
        BNS+BNS & $0.13^{+0.71}_{-0.12}$ & $26.0^{+140.0}_{-25.0}$ & $14000.0^{+22000.0}_{-11000.0}$ \\ \vspace{1mm}
        BNS+BBH & $0.11^{+0.18}_{-0.07}$ & $22.0^{+36.0}_{-13.0}$ & $13000.0^{+8300.0}_{-4800.0}$ \\
        \bottomrule\bottomrule
    \end{tabular}
    \caption{Estimated number of signals occurring for different CBC overlap configurations in a year's observations of the Einstein Telescope. The first column, $N_{events}$ is an estimate of the total number of single signal events, of that kind, that ET would expect to see in the year's observation. For the BNS+BBH row this is the expected number of BBHs, rather than the number of BNSs. The other columns represent the predicted number of overlaps of this kind in each bias region.}
    \label{tab:rates_table}
\end{table*}

Table \ref{tab:rates_table} shows estimated numbers of events falling within each of the bias regions, defined above, over a year's observation of ET \cite{punturo2009einstein}. These numbers were estimated by calculating the visible volume for a variety of different mass CBCs as described in \cite{relton2021parameter}, using the \texttt{inspiral-range} \texttt{PYTHON} package ~\cite{chen2021distance, gwinc-inspiralrange}, and a PSD for ET \cite{ETPSD}. The viewing time for each signal, in the negligible bias region, was estimated from $\SI{1}{Hz}$ to the time of merger. Signals in the strong and weak bias regions were fixed to the durations described above, merger time separations of $\SI{0.1}{s}$ and $\SI{2}{s}$, respectively, for BBH signals. For a more detailed explanation of this method see Section II-B of \cite{relton2021parameter}. 

From the numbers in Table \ref{tab:rates_table} it is clear that, despite the strong and weak regions being the most cause for concern, only a very small fraction of events will fall in this region. By far the dominant case is the negligible bias region, into which all signals in these detectors should fall.

%----------------------------------------------------------
%---------------------- BBH Section -----------------------
%----------------------------------------------------------
\subsection{Injection studies} \label{cbcresults}
\subsubsection{BBH+BBH overlaps} \label{bbhresults}
\begin{table*}[ht]
    \centering
    \begin{tabular}{lllllll}
        \toprule\toprule
        \multicolumn{2}{c}{\multirow{2}{*}{\textbf{BBH+BBH}}} & \multirow{2}{*}{Injected} & \multicolumn{2}{c}{PyCBC} & \multicolumn{2}{c}{cWB} \\
        {} & {} & {} & SINGLES & PAIRS & SINGLES & PAIRS \\
        \midrule
        \multirow{2}{*}{Total} & Counts & $13172$ & $10454$ & $9818$ & $6885$ &    $5634$ \\
        {} & Percentage &  -  &  $79.37\%$ & $74.54\%$ & $52.27\%$ & $42.77\%$ \\
        \midrule
        \multirow{2}{*}{FAR$<\SI{1}{yr^{-1}}$} & Counts & $13172$ & $8436$ & $7947$ & $6883$ & $5310$ \\
        {} & Percentage &  - & $64.04\%$ & $60.33\%$ & $52.25\%$ & $40.31\%$ \\
        \bottomrule\bottomrule
    \end{tabular}
    \caption{Injected and recovered individual overlapping signals in different injection sets and search pipelines. Here the two signals in a pairing are both BBH mergers. The SINGLES column here is the union of the results from both $\mathrm{SINGLES}_{\mathrm{A}}$ and $\mathrm{SINGLES}_{\mathrm{B}}$ data sets. The FAR threshold of $<\SI{1}{yr^{-1}}$ means that, in a years observation, fewer than one event of this kind will occur due to statistical fluctuations in the noise. This is a fairly typical cut for assuring an event is astrophysical. The total column is for all events, without the threshold.}
    \label{tab:BBH_results}
\end{table*}

Table \ref{tab:BBH_results} contains values of the number of injected and recovered signals in each injection set in both \texttt{PyCBC}) and \texttt{cWB}. As can be seen, in \texttt{PyCBC}), the vast majority, close to $80\%$, of injected non-overlapping signals were recovered by any trigger. The other $20\%$ were largely missed due to the signals having low network SNRs, $\lesssim 12$, or being in less sensitive sky locations for the network. However, another $15\%$ are removed once a FAR threshold of $<\SI{1}{yr^{-1}}$ is applied, these could be true astrophysical signals, but are rejected due to poor significance. This is an artefact of the injected distribution to provide a reasonable number of detectable events.

As expected \cite{abbott2021population}, \texttt{cWB} is less sensitive, with respect to modelled searches, for this range of masses. The missed signals here are largely low mass CBCs, $\mathcal{M} \lesssim 15 \, M_{\odot}$, in which the majority of the SNR comes from the inspiral, which is difficult for \texttt{cWB} to recover. As expected, if the injections are present in the template bank of the matched filter search, then the unmodelled method will always be less optimal than the matched filter searches. This is the case here since our injections have source parameters which are well described by the template bank.

\begin{figure*}[t]
    \centering
    \includegraphics[width=\linewidth]{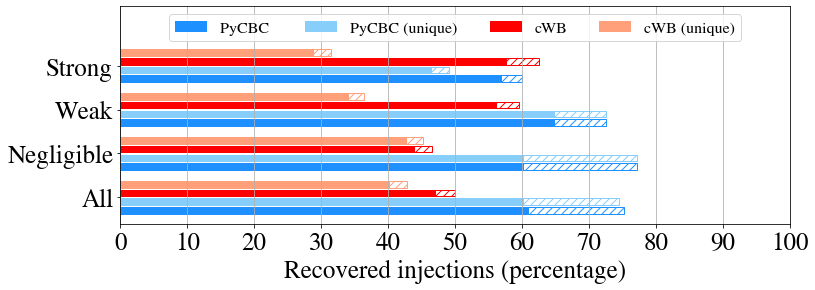}
    \caption{Bar charts for the found injections in both pipelines in the different overlap bias regions. Each region has four bars, split into two for each pipeline. These show the percentage injections in which the pipeline found a trigger. The ``unique" column shows the percentage of unique triggers, i.e. if both signals are found by the same trigger, then they are only counted once. The shaded segments show injections recovered at a FAR threshold of $<\SI{1}{yr^{-1}}$. The data here is for the BBH+BBH injection sets.}
    \label{fig:BBH_bar}
\end{figure*}

Figure \ref{fig:BBH_bar} shows bar charts for the percentage of injections that were found by a trigger in each pipeline in the PAIRS injection set. As expected the pipelines perform reasonably well in the negligible bias region, with the majority of signals found in \texttt{PyCBC}). For both pipelines, in this region, the number of injections found matches the findings in the SINGLES injections. There is some decrease in the fraction successfully caught when moving to the weak and strong bias regions.

Inside the strong bias region for \texttt{PyCBC}), and some of the weak region for \texttt{cWB}, some of the injections are both found in the same trigger. This means that they are counted twice. To show this we have added extra bars for each search in which we have removed duplicated triggers here. This is an effect of the criteria we have applied to decide if an injection is found. See Section \ref{findingcriteria} for more details.

The strong bias region here shows a functional problem with the pipelines for signals in this region. These numbers are slightly lower than they theoretically should be, due to the clustering of triggers in the pipeline. In \texttt{PyCBC}), as templates are matched to the data, numerous triggers are recorded for each signal as multiple templates may match the signal to differing levels of significance. To avoid all of these triggers being returned for a single signal, a clustering window is set. This window, set to $\pm\SI{1}{s}$ in our study, clusters these triggers and returns a single trigger of highest significance.

\begin{figure}[t]
    \centering
    \includegraphics[width=\linewidth]{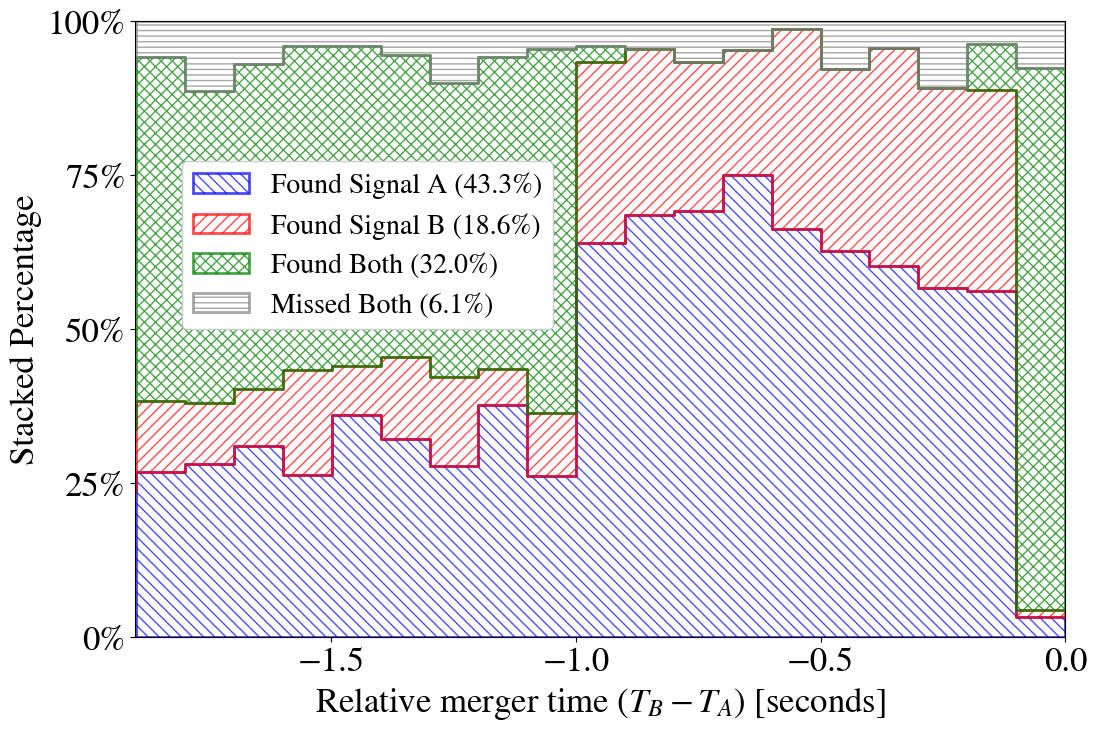}
    \caption{A stack plot showing the distribution of how injected signals were found by relative merger time in the \texttt{PyCBC}) BBH+BBH PAIRS injection set. Blue and red show the pairings in which only one signal was found, Signal A or Signal B respectively. Green shows the the pairings in which both signals were found, grey for entirely missed pairs. This plot covers both the strong and weak bias regions. The time convention is $T_B - T_A$, this is always less than $0$ due to the convention of drawing Signal B's merger within the observable duration of Signal A. The percentages in the caption refer to the fraction of pairings in this region, $|\Delta t_c| \leq \SI{2}{s}$, that fall in each category.}
    \label{fig:merger_time_stack}
\end{figure}

In the case of overlapping signals within this $\pm\SI{1}{s}$ merger time separation window, only the most significant signal is returned. As such, in the strong bias region and some of the weak bias region, many of the signals are missed due to only the most significant being returned. In our study the triggered signal is usually Signal A. Due to our method of drawing of the merger time, Signal B always merges before Signal A. Therefore, in a matched-filter search, Signal A is favoured as it is still providing power in the data at this time. This is not necessarily the case with unmodelled searches, such as \texttt{cWB}, see Section \ref{cwb_separation} for further details. An example of this can be seen in Figure \ref{fig:merger_time_stack}. 

\begin{figure}[t]
    \centering
    \includegraphics[width=\linewidth]{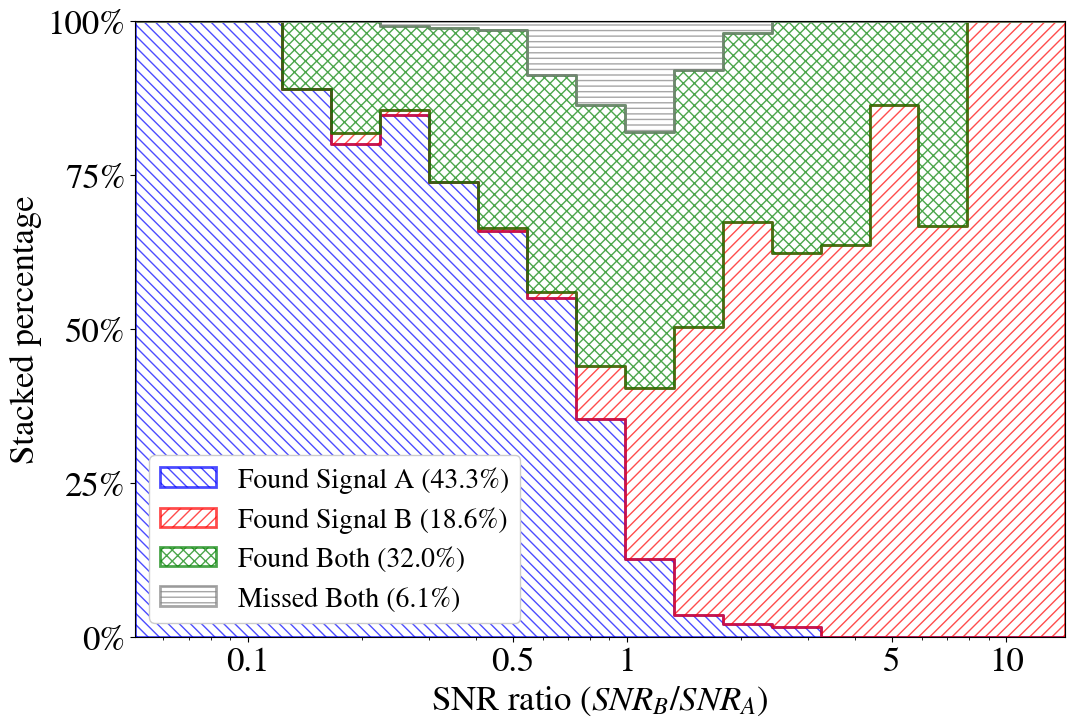}
    \caption{A stack plot showing the distribution of how pairings were found as a function of SNR ratio between the two signals. The ratio is defined to be $\text{SNR}\textsubscript{B} / \text{SNR}\textsubscript{A}$. Colours are consistent with those in Figure \ref{fig:merger_time_stack}. Similarly, the data comes from pairings in the strong and weak bias regions of the \texttt{PyCBC}) BBH+BBH PAIRS injection set. The percentages in the caption refer to the fraction of pairings in this region, $|\Delta t_c| \leq \SI{2}{s}$, that fall in each category.}
    \label{fig:SNR_stack}
\end{figure}

Both found triggers are only in the region outside the clustering window of \texttt{PyCBC}), $|\Delta t_c| \leq \SI{1}{s}$\footnote{Three pairings have both signals found inside the clustering window. These pairings have merger time separations of approximately $-\SI{0.999}{s}$ and, in some detectors have separations $|\Delta t_c| > \SI{1}{s}$}. In the region $|\Delta t_c| \leq \SI{1}{s}$, Signal A is clearly favoured, as it is the later merging signal. Pairings in this region that are found with Signal B are largely either very close in merger time or have $\text{SNR}\textsubscript{B} > \text{SNR}\textsubscript{A}$. This is shown in Figure \ref{fig:SNR_stack}, where the distribution of SNR ratio between signals in found pairings is shown. In the region $<\SI{1}{s}$ here Signal A is louder, and therefore favoured, and in the region $>\SI{1}{s}$ Signal B is louder and more likely to be found. Missed pairings here have generally even SNR ratios due to them both having poor SNRs. These regions are artefacts of the pipeline settings and as such remain the same for the longer signals in the BNS+BNS and BNS+BBH runs outlined in Sections \ref{bnsresults} and \ref{bothresults}.

For Figures \ref{fig:merger_time_stack} and \ref{fig:SNR_stack}, we have only included \texttt{PyCBC}) results. This is due to the large time window criteria we applied to count found \texttt{cWB} injections In this region, most \texttt{cWB} pairings will be single signal found. However, most signals with such close mergers will only be found as a single trigger in \texttt{cWB}. \texttt{cWB} internally applies time windows on regions of excess power, under the assumption that it would only ever be a single signal. In wider regions, such as the negligible bias region, the search returns separate triggers for each component signal in the pairing. This is akin to the clustering window of \texttt{PyCBC}), however, when \texttt{PyCBC}) finds both injections by a single trigger it will return a template for a single signal under the assumption that it is a single signal, \texttt{cWB} returns the coherent power of one, or potentially both signals, see Section \ref{Identification}.

There are some signals in the negligible bias region that are found in the single signal injection sets, but missed in the overlapping runs. Generally, these signals have relative merger times just beyond the two second negligible bias region and have fairly uneven SNR-ratios. This follows the findings of previous parameter estimation studies. In those studies, if one signal is much louder than the other then samples are recovered matching the parameters of the louder signal. In the case of match filter searches, the template bank search will recover templates closer to the louder signal, the remaining power from the weaker signal will be rejected as either noise, or excess power in the inspiral of the louder signal.

The drop in efficiency due to the presence of time-overlapping signals can be estimated by comparing the efficiency of the overlapping and non-overlapping signals. For \texttt{PyCBC}) this drop off is approximately $4\%$ across the entire run. In \texttt{cWB} this drop in efficiency is approximately $12\%$. The differences in pipeline efficiency means that a direct comparison is not a simple matter, we also report the relative efficiency drops as $6\%$ and $23\%$ for \texttt{PyCBC}) and \texttt{cWB} respectively.

The errors on all efficiencies are less than $0.01\%$ and as such were not included. All quoted efficiencies are for events found with a FAR threshold of $< 1$ per year.

Outside the clustering window of \texttt{PyCBC}), this fall in efficiency is approximately $1\%$, while inside the clustering window the fall in efficiency is $26\%$ due to the majority of paired signals being found by a single trigger, or not at all. These regions are not directly comparable for \texttt{cWB}, in these cases it will find both signals and return them as one trigger, however, similar numbers would be $9\%$ and $32\%$ for outside and inside the $|\Delta t_c| = \SI{1}{s}$ boundary.

%----------------------------------------------------------
%---------------------- BNS Section -----------------------
%----------------------------------------------------------
\subsubsection{BNS+BNS overlaps} \label{bnsresults}
\begin{table*}[ht]
    \centering
    \begin{tabular}{lllllll}
        \toprule\toprule
        \multicolumn{2}{c}{\multirow{2}{*}{\textbf{BNS+BNS}}} & \multirow{2}{*}{Injected} & \multicolumn{2}{c}{PyCBC} & \multicolumn{2}{c}{cWB} \\
        {} & {} & {} & SINGLES &   PAIRS & SINGLES &   PAIRS \\
        \midrule
        \multirow{2}{*}{Total} & Counts & $2212$ & $2042$ & $1677$ & $554$ & $467$ \\
        {} & Percentage &  -  & $92.31\%$ & $75.81\%$ & $25.05\%$ & $21.11\%$ \\
        \midrule
        \multirow{2}{*}{FAR$<\SI{1}{yr^{-1}}$} & Counts & $2212$ & $1854$ & $1544$ & $550$ & $461$ \\
        {} & Percentage &  - &  $83.82\%$ & $69.80\%$ & $24.86\%$ & $20.84\%$ \\
        \bottomrule\bottomrule
    \end{tabular}
    \caption{Injected and recovered individual overlapping signals in different injection sets and search pipelines. The SINGLES column here is the union of the results from both $\mathrm{SINGLES}_{\mathrm{A}}$ and $\mathrm{SINGLES}_{\mathrm{B}}$ data sets. Here the two signals in a pairing are both BNSs. It should be noted that these values are not directly comparable to those in Table \ref{tab:BBH_results}, as the luminosity distances were arbitrarily set for BBH and BNS injections such that most were visible in the detector. As such we have a slightly higher proportion of BNS signals recovered in \texttt{PyCBC} than we did for BBH signals.}
    \label{tab:BNS_results}
\end{table*}

\begin{figure*}[t]
    \centering
    \includegraphics[width=\linewidth]{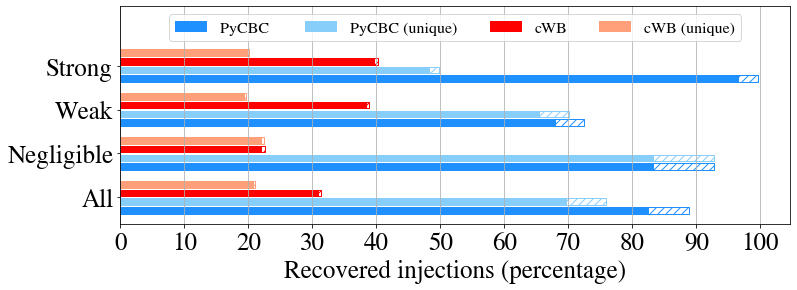}
    \caption{Bar charts for the found injections in both pipelines in the different overlap bias regions. Each region has four bars, split into two for each pipeline. These show the percentage injections in which the pipeline found a trigger. The ``unique" column shows the percentage of unique triggers, i.e. if both signals are found by the same trigger, then they are only counted once. The shaded segments show injections recovered at a FAR threshold of $< \SI{1}{yr^{-1}}$. The data here is for the BNS+BNS injection sets.}
    \label{fig:BNS_bar}
\end{figure*}

Table \ref{tab:BNS_results} shows the results for the BNS+BNS overlap injection set. There are fewer signals in these injection sets as a higher number of signals would have caused $N_{signals}>2$ overlaps, which we do not consider in this study. As expected, \texttt{cWB} finds a lower percentage of the injections, as it is not designed to find longer, inspiral dominated signals such as binary neutron stars. This can be seen in Figure \ref{fig:BNS_bar} where \texttt{cWB} recovers only about $20\%$ of injections with unique triggers. \texttt{PyCBC}) has a slightly higher efficiency here than in the BBH+BBH run, however, this is most likely a consequence of the injected luminosity distances of the signals rather than the design of the search pipeline.

Due to these long durations and the uniform drawing of merger time separation, the signals were drawn such that approximately half fell into the negligible bias region with a further quarter falling in each of the weak and strong bias regions. As the strong bias region of BNS+BNS events is, as predicted by parameter estimation studies, $|\Delta t_{c}| \leq \SI{0.01}{s}$, a very high percentage of injected pairs fall inside the \texttt{PyCBC}) clustering window so are both found by the same trigger. This is apparent in the strong region of Figure \ref{fig:BNS_bar} where the unique bar is half that of the non-unique found.

The efficiency drops between overlapping and non-overlapping BNS signals is $14\%$ for \texttt{PyCBC}) and $4\%$ for \texttt{cWB}. The \texttt{cWB} values drop is small, compared to \texttt{PyCBC}), as the majority of BNS signals found by \texttt{cWB} are very significant regardless of overlap. The relative drop in these efficiencies are $17\%$ and $16\%$ respectively.

%----------------------------------------------------------
%---------------------- BOTH Section ----------------------
%----------------------------------------------------------
\subsubsection{BNS+BBH overlaps} \label{bothresults}
\begin{table*}[ht]
    \centering
    \begin{tabular}{lllllll}
        \toprule\toprule
        \multicolumn{2}{c}{\multirow{2}{*}{\textbf{BNS+BBH}}} & \multirow{2}{*}{Injected} & \multicolumn{2}{c}{PyCBC} & \multicolumn{2}{c}{cWB} \\
        {} & {} & {} & SINGLES & PAIRS & SINGLES & PAIRS \\
        \midrule
        \multirow{2}{*}{Total} & Counts & $2400$ & $2086$ & $1713$ & $928$ & $706$ \\
        {} & Percentage &  -  & $86.92\%$ & $71.38\%$ & $38.67\%$ & $29.42\%$ \\
        \midrule
        \multirow{2}{*}{FAR$<\SI{1}{yr^{-1}}$} & {Counts} & $2400$ & $1782$ & $1519$ & $922$ & $673$ \\
        {} & Percentage &  - & $74.25\%$ & $63.29\%$ & $38.42\%$ & $28.04\%$ \\
        \bottomrule\bottomrule
    \end{tabular}
    \caption{Injected and recovered individual overlapping signals in different injection sets and search pipelines. The SINGLES column here is the union of the results from both $\mathrm{SINGLES}_{\mathrm{A}}$ and $\mathrm{SINGLES}_{\mathrm{B}}$ data sets. Here the two signals in a pairing are a BNS and a BBH. It should be noted that the comparison between \texttt{PYCBC} and \texttt{cWB} for the SINGLES runs is challenging as cWB will perform significantly differently for the BNS SINGLES run. See Table \ref{tab:BNS_results} for a more precise comparison of single signal runs for these events.}
    \label{tab:BOTH_results}
\end{table*}

As described in Section \ref{bnsresults} \texttt{cWB} is not tuned to low mass, inspiral dominated signals, like BNSs. As such it does not perform well for the BNS portion of the SINGLES injection sets. The efficiencies of the pipeline for these injection sets highlights this. For \texttt{cWB}, the $\mathrm{SINGLES}_{\mathrm{A}}$ run recovered $25\%$ of injections, while the $\mathrm{SINGLES}_{\mathrm{B}}$ run recovered $52\%$, with $86\%$ and $82\%$ being the comparable numbers for \texttt{PyCBC}).

\begin{figure*}[t]
    \centering
    \includegraphics[width=\linewidth]{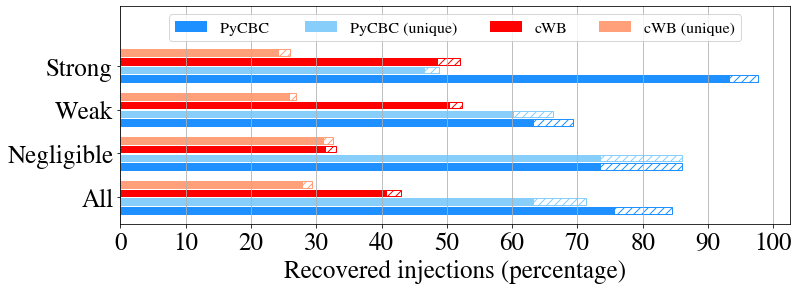}
    \caption{Bar charts for the found injections in both pipelines in the different overlap bias regions. Each region has four bars, split into two for each pipeline. These show the percentage injections in which the pipeline found a trigger. The ``unique" column shows the percentage of unique triggers, i.e. if both signals are found by the same trigger, then they are only counted once. The shaded segments show injections recovered at a FAR threshold of $< \SI{1}{yr^{-1}}$. The data here is for the BNS+BBH injection sets.}
    \label{fig:BOTH_bar}
\end{figure*}

Table \ref{bothresults} shows a significant falloff between the SINGLES and PAIRS injection sets, compared to the BBH+BBH run. This, as in Section \ref{bnsresults}, is in large part due to a large percentage of strong bias pairs falling inside the clustering windows and being found by the same trigger. This can be seen in Figure \ref{fig:BOTH_bar} where, as in Figure \ref{fig:BNS_bar} the strong bias region sees a significant falloff from any to unique found injections.

Here \texttt{PyCBC}) performs better for BNS+BBH overlaps than for BNS+BNS overlaps. The same cannot be said for \texttt{cWB}, which would appear to perform more consistently in the BNS+BNS run. Although this may be difficult to conclude due to the low efficiency of that run.

The efficiency drop between overlapping and non-overlapping BNS signals is $11\%$ for \texttt{PyCBC}) and $10\%$ for \texttt{cWB}. Care should be taken here in comparisons between the pipelines, as \texttt{PyCBC}) has similar sensitivities to BBH and BNS signals, whereas \texttt{cWB} will favour higher mass signals and is therefore more sensitive to one half of the non-overlapping signals. The relative drops in efficiency are $15\%$ and $26\%$ respectively.

\subsection{Accuracy of recovered triggers} \label{trigger_accuracy}

\begin{figure}[t]
    \centering
    \includegraphics[width=\linewidth]{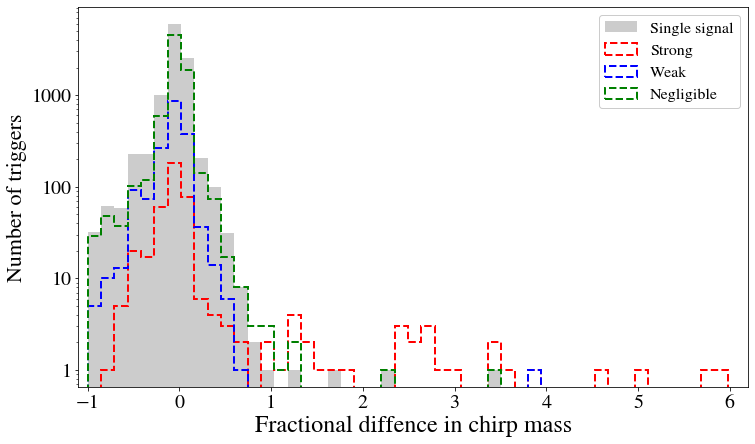}
    \caption{Distribution, across the overlap regions, of the
    fractional difference between the recovered and injected chirp masses for \texttt{PyCBC}) triggers in the PAIRS injection set. Also shown, in grey, is the same distribution for recovered and injected in the SINGLES injection set. A perfect pipeline, with an infinitely finely gridded template bank, would find injections with a delta spike on zero in this plot. For direct comparison across the overlap regions, we have included this plot as a function of regional density in Appendix \ref{appendix1}.}
    \label{fig:delta_Mc}
\end{figure}

Figure \ref{fig:delta_Mc} contains histograms of the fractional chirp mass difference between the triggers recovered, by \texttt{PyCBC}), in the PAIRS injection set and the true injected values. It also includes a similar distribution for the SINGLES injection set, shown in grey.

In most cases the recovered trigger in the PAIRS injections match what is found in the single signal injections to a reasonable level, $|\Delta \mathcal{M}| < 1 \, M_\odot$. This can be seen by comparing the distribution of negligible bias region injections to the SINGLES distribution. The weak and negligible regions match this distribution fairly well, with few outliers. However, the strong bias region skews towards a high fractional difference, up to $\sim6$ times the value found in SINGLES. Therefore, we can say that signal pairings in the strong bias region are affected in a similar way in matched filter searches to matched filter based parameter estimation.

The skew of this distribution shows that incorrectly found signals tend to be found with higher chirp mass templates than the injected signal. As this is also true for the inaccurately found single signals, this is a feature of how we calculate the fractional difference. The recovered chirp mass can never be less than -1 as for a trigger to be recovered it must have a positive chirp mass. Triggers with much larger values of fractional chirp mass are likely extreme due to the lower density of templates at the high mass tail of the template bank.

%----------------------------------------------------------
%-------------------- Identification ----------------------
%----------------------------------------------------------
\section{Overlap Identification and Separation} \label{Identification}
\begin{figure*}[t]
    \centering
    \includegraphics[width=\linewidth]{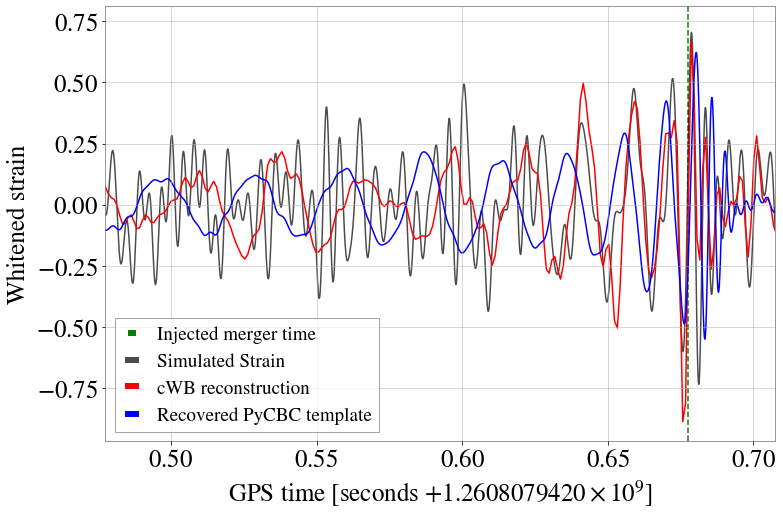}
    \caption{A plot of whitened strain, in the LIGO-Hanford detector, for a single BBH injection. The blue lines shows the matched template found by \texttt{PyCBC}) in the $\mathrm{SINGLES}_{\mathrm{A}}$ injection set. The red line is the \texttt{cWB} reconstructed waveform from the same injection set. The green, dashed, vertical line indicates the injected merger time.}
    \label{fig:waveform_comparison_SINGLE}
\end{figure*}

\begin{figure*}[t]
    \centering
    \includegraphics[width=\linewidth]{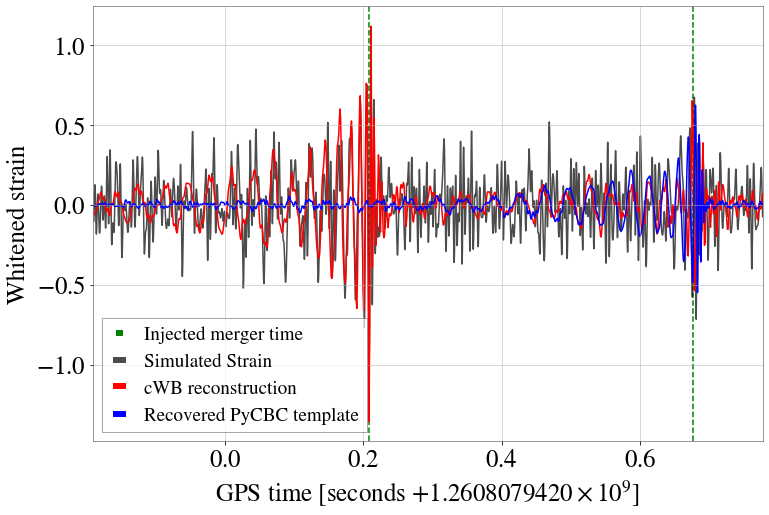}
    \caption{A plot of whitened strain, in the LIGO-Hanford detector, for a BBH+BBH injection. The blue line shows the matched templates found by \texttt{PyCBC}) in the PAIRS injection set. The red line is the \texttt{cWB} reconstructed waveform from the PAIRS injection set. The green, dashed, vertical lines indicates the injected merger times.}
    \label{fig:waveform_comparison_PAIR}
\end{figure*}

Figure \ref{fig:waveform_comparison_SINGLE} shows a comparison of the whitened strain of an injected BBH single signal in simulated Hanford detector strain. This is accompanied by a whitened version of the template found in the \texttt{PyCBC}) search and the reconstructed waveform, produced by \texttt{cWB}. Here, outside the merger, the template found by \texttt{PyCBC}) does not match the signal as well as the \texttt{cWB} reconstruction. This is most noticeable at lower frequency, although the mergers are broadly similar. This is due to the coarseness of the gridding in the \texttt{PyCBC}) template bank and the lack of bank sensitivity to parameters such as spin and phase.

Figure \ref{fig:waveform_comparison_PAIR} shows an equivalent plot involving the Signal A represented in Figure \ref{fig:waveform_comparison_SINGLE} overlapping with a Signal B. It can be seen that Signal B, merging first, is found in \texttt{cWB} but not in \texttt{PyCBC}). Both signals are significant, with network SNRs of 26 and 27. Despite this, due to the close mergers of the signals and clustering, the template for Signal A is returned in the \texttt{PyCBC}) PAIRS run due to its later merger time. The Signal B from this pairing was found by both \texttt{PyCBC}) and \texttt{cWB}, as an individual signal, in the $\mathrm{SINGLES}_{\mathrm{B}}$ injection set.

The phase of the \texttt{cWB} reconstruction for Signal A in Figure \ref{fig:waveform_comparison_PAIR} is different compared to that showing in Figure \ref{fig:waveform_comparison_SINGLE} as the algorithm is trying to fit both signals to the same, coherent sky location. Indeed, as shown in Figures \ref{fig:SINGLES_cWB} and \ref{fig:PAIRS_cWB}, in that case the likelihood maximisation is considerably affected by Signal B, and so Signal A's reconstruction is not optimal. This is discussed in detail in Section \ref{cwb_separation}.

\subsection{Separation through unmodelled algorithms} \label{cwb_separation}

\begin{figure*}[t]
    \centering
    \includegraphics[width=\linewidth]{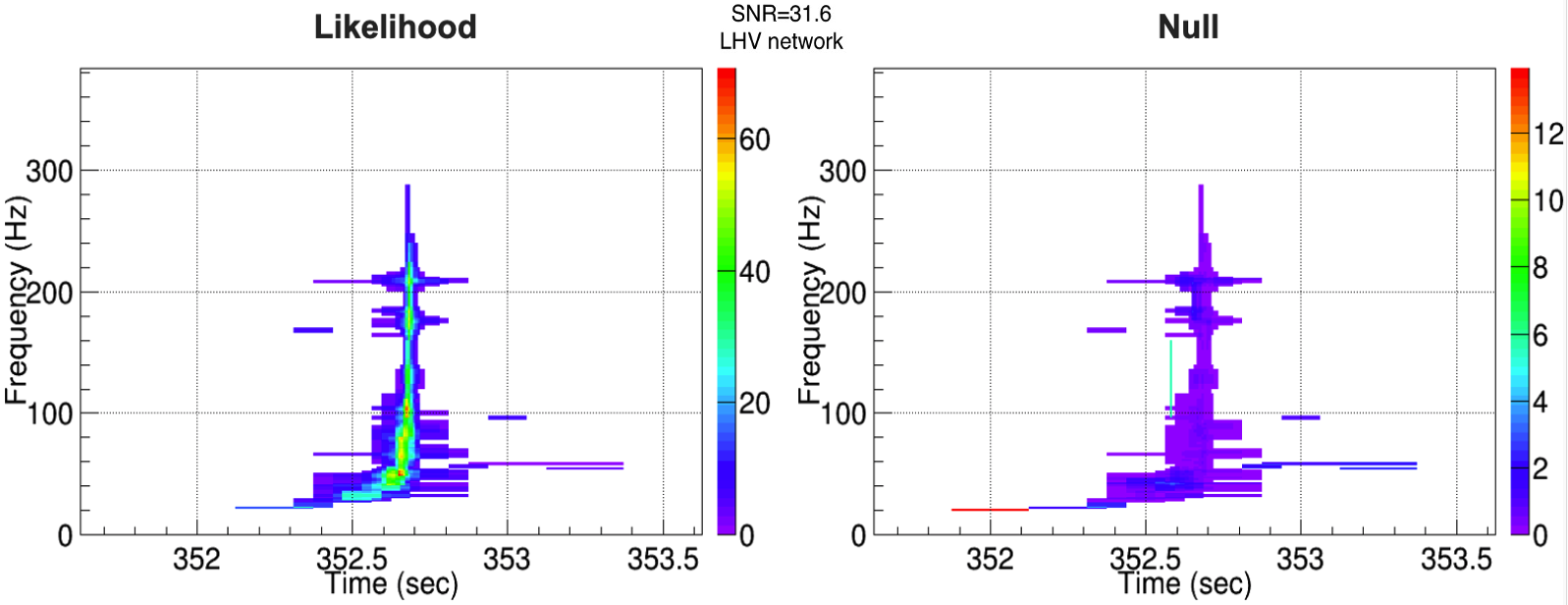}
    \caption{A spectrogram of likelihood and null energy for a single signal event. The likelihood contains almost entirely signal energy, while the null almost entirely noise. The data is the same as for the Single A signal in Figure \ref{fig:waveform_comparison_SINGLE}}
    \label{fig:SINGLES_cWB}
\end{figure*}

The \texttt{cWB} framework, as described in Section \ref{searchcwb}, is designed to analyse a single signal. The pipeline finds regions of excess power and then maximises the likelihood with respect to one source location, regardless of the number of present astrophysical signals. In the standard case of a single signal trigger, the pipeline maximises the likelihood with respect to that signal, and so almost all its energy is placed in the likelihood, while the null is almost entirely noise, as shown in Figure \ref{fig:SINGLES_cWB}.

For the case of two overlapping signals, the likelihood is then maximised considering both signals, generally with one favoured over the other, considering this signal as the ``primary" and the other as the ``secondary". In this case, the likelihood is largely maximised with respect to the primary signal, though often with some contamination from the secondary signal, depending on its energy. This means that the primary signal energy is almost entirely found in the likelihood. The energy of the secondary signal is divided between likelihood and null depending on its source location. If this is considerably different with respect to that of the primary signal, then a considerable amount of the energy of the secondary signal will remain in the null. This is similar to the situation in Figures \ref{fig:waveform_comparison_SINGLE} and \ref{fig:waveform_comparison_PAIR}.

\begin{figure*}[t]
    \centering
    \includegraphics[width=\linewidth]{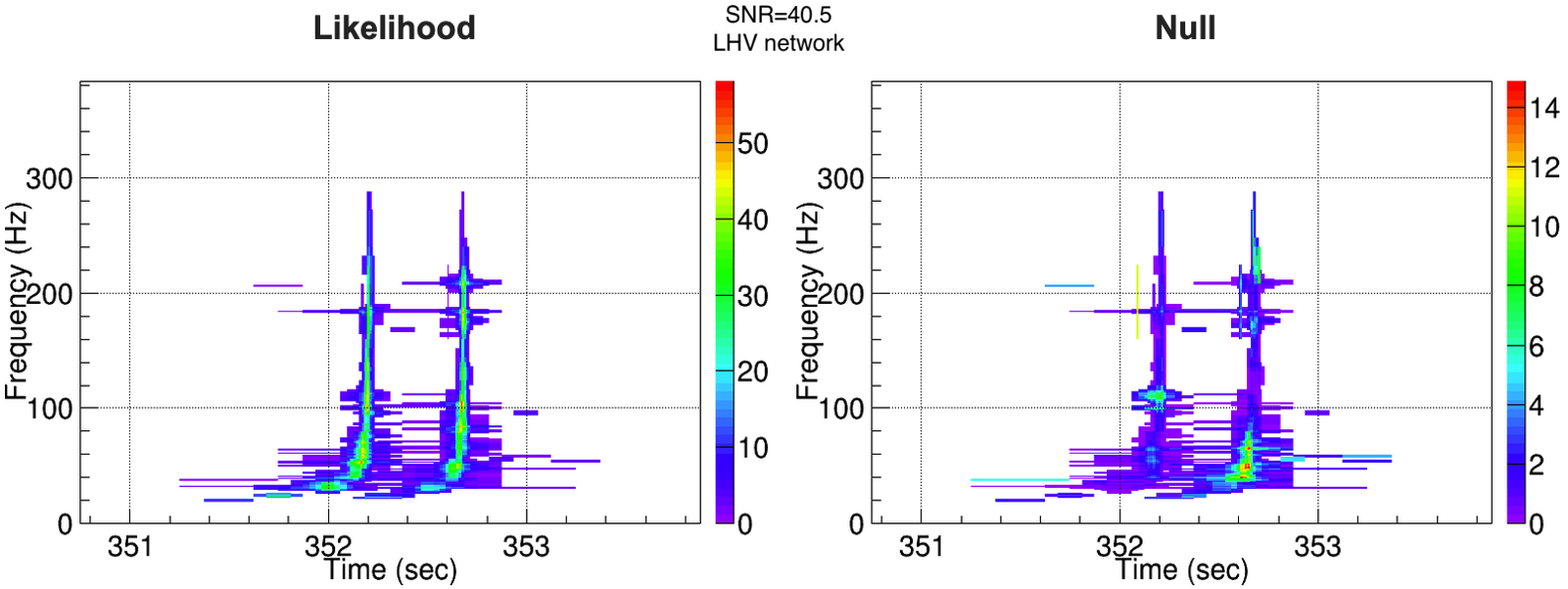}
    \caption{A spectrogram of likelihood and null energy for two overlapping signals. The likelihood has been maximised with respect to the primary signal, so a relevant fraction of the energy associated with the secondary one remains in the null. The data is the same as for the signals in Figure \ref{fig:waveform_comparison_PAIR}}.
    \label{fig:PAIRS_cWB}
\end{figure*}

Figure \ref{fig:SINGLES_cWB} shows the time-frequency representation of the likelihood and null energies of Signal A, in which likelihood has been properly maximised. The signal is almost entirely caught in the likelihood. The remaining energy in the null is likely excess noise. Figure \ref{fig:PAIRS_cWB} represents the same signal in an overlapping pairing, as in Figure \ref{fig:waveform_comparison_PAIR}. In this case, the likelihood is maximised mainly with respect to Signal B, here the primary signal. The null contains the secondary, here Signal A, and is increased with respect to the single detection. The likelihood has been maximised with respect to a different sky location, and both the null and likelihood contain a non-negligible contribution from each signal.

This contribution of the signal to the residual noise energy increases the penalty, lowering the correlation coefficient. For this reason, triggers with overlapping signals are penalised when applying post-production cuts, as it can be seen in detection efficiencies reported in Tables \ref{tab:BBH_results}, \ref{tab:BNS_results} and \ref{tab:BOTH_results} and in Figures \ref{fig:BBH_bar}, \ref{fig:BNS_bar} and \ref{fig:BOTH_bar}.

These results suggest that, to produce an optimal analysis of overlapping signals, a future development of \texttt{cWB} should allow for the estimation of multiple likelihoods. If a TF map suggests the presence of two overlapping signals, then the information about the way to analyse it can be obtained firstly analysing the data with a likelihood following the primary signal. If the null then contains an indication for another signal, then it could be used to select the pixels related to this signal and try to maximise the likelihood with respect to them only. It is important to remark that the TF representation is able to disentangle, by itself, the two overlapping signals only if these cover different TF pixels. In this case, the separate maximisation of two different likelihoods, via two different sets of pixels, would give the signal reconstruction as optimally as it is currently for single signal triggers. It should be noted that in the unlikely case in which the two signals come from almost the same source location the two likelihoods would be almost the same.

If a pixel has contributions from both signals, then the TF separation this is not possible. In that case there is still the possibility to exploit the likelihood-null energy distribution for the involved overlapping pixels, if the signals come from different sky locations. When the two signals cover the same pixel the likelihood-null disentangling would be non-optimal as some fraction of the secondary signal will fall into the likelihood of the primary, as shown in Figure \ref{fig:PAIRS_cWB}. This implies that we may not be able to fully recover the secondary signal from the null. If the signals come from almost the same source location, then there is no way to disentangle them in an unmodelled way, at least in those TF pixels covered by both signals: the likelihood-null approach fails as well because in that case also the null is almost the same for both signals. All these considerations should be examined in future studies.

\subsection{Separation through matched filter algorithms} \label{mf_separation}
When searching for signals in data, matched filter searches, such as \texttt{PyCBC}), will match a template waveform to the data at time intervals in order to create an SNR time series of the data. This time series is a map of the SNR of that template with the data. When one of these templates closely matches a signal, the time series will peak.

\begin{figure}[t]
    \centering
    \includegraphics[width=\linewidth]{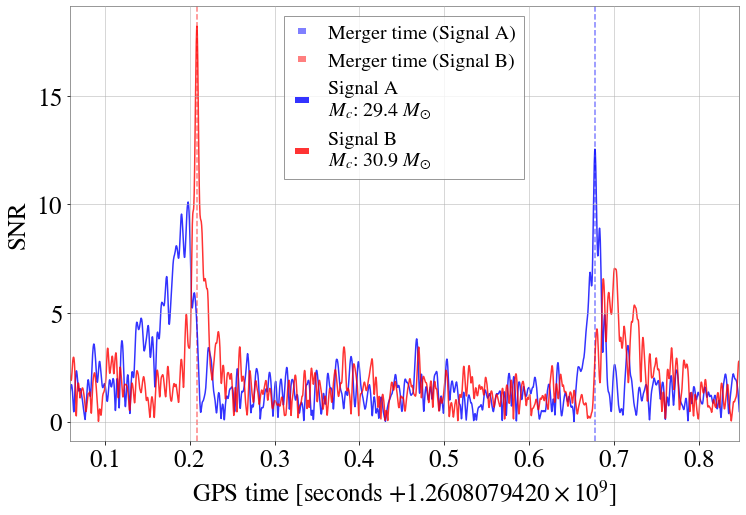}
    \caption{A plot of SNR time series for two perfect templates against simulated data containing both signals in the LIGO-Hanford detector. The blue line shows the SNR of the Signal A template against the data, the red line shows the equivalent for the Signal B template. The blue and red, dashed, vertical lines show the injected merger times of Signal A and B respectively. The signals are the same as those shown in Figure \ref{fig:waveform_comparison_PAIR}. The peak SNRs here are smaller than those given in Section \ref{Identification} as they are single detector SNRs, not network SNRs.}
    \label{fig:SNR_TS}
\end{figure}

In a situation in which there are two signals in the data, this time series should peak twice, once for each signal in the data. Figure \ref{fig:SNR_TS} shows this for two perfect templates against data with two signals injected. The blue line, for Signal A's template, shows a small peak around the merger of Signal B, where the match is not perfect. It then peaks again, cleanly and more significantly, around the merger of Signal A. The reverse is shown for a perfect Signal B template in red.

The signals and templates used in Figure \ref{fig:SNR_TS} have very similar chirp masses, and as such match reasonably well with each other's templates. Conversely, if the signals have very different chirp masses, then this plot will only show a single peak for each signal. Although, there will likely be some non-Gaussianity in the region of the non-ideal signals merger time.

\texttt{PyCBC})'s clustering routines, as described in Section \ref{findingcriteria}, will ignore these features and only return the maximum likelihood signal, in this case Signal A. It should be possible to modify the clustering routines to check for multiple peaks in a single template, or for single peaks in multiple templates. These results can then be used to identify reasonable parameter ranges for each signal in the overlap. This would allow for a reliable starting point for multi-signal parameter estimation \cite{antonelli2021noisy}.

%----------------------------------------------------------
%---------------------- Discussion ------------------------
%----------------------------------------------------------
\section{Discussion} \label{discussion}
We showed here that both modelled, and unmodelled searches can detect overlapping CBC signals in the negligible bias region of $|\Delta t_{c}| > \SI{2}{s}$. Within the weak bias region, so long as the signals are separated by more than the matched filter clustering window, here $|\Delta t_{c}| > \SI{1}{s}$, the matched filter search should recover both signals, provided one is not much louder than the other. We also show that in the narrower weak and strong bias regions both pipelines can successfully recover one or both signals in the overlap most of the time. 

Inside the clustering window, matched filter pipelines do struggle to successfully recover both signals due to internal methods designed to reduce the number of false triggers returned. However, we believe that these searches can be modified to find templates that best match each signal in the pairing. This could provide reasonable best guesses for multi-signal parameter estimation.

We showed that unmodelled searches can successfully recover both signals in the pairing, both as a single trigger when the mergers are close, and as separate triggers when far apart. \texttt{cWB}, when recovering signals via coherent sky location, is able to disentangle overlapping signals, so long as they cover different frequencies in the TF map. In the case in which the signals are in the same TF pixels, the secondary signal should be at least partially separable via the null energy map, if coming from a different sky location than the first. This should be another indicator of the presence of a secondary signal. It is possible that the second signal's sky location may be recoverable leading to further first guess parameters for multi-signal parameter estimation.

We believe that, with modifications, the combination of these two forms of CBC search pipelines should provide a reliable method for the detection, identification and initial separation of time-overlapping CBCs.

%----------------------------------------------------------
%------------------- Acknowledgements ---------------------
%----------------------------------------------------------
\section{Acknowledgements} \label{acknowledgements}

We would like to thank Tim Dietrich, Stephen Fairhurst, Jonathan Thompson, Alex Nitz, Edoardo Milotti, and Giovanni Prodi for useful discussions. We would also like to thank the anonymous reviewer. We are grateful for computational resources provided by Cardiff University and funded by an STFC grant supporting UK Involvement in the Operation of Advanced LIGO. The authors are grateful for computational resources provided by the LIGO Lab (CIT) and supported by Natiwith the network of detectors by correlating the dataonal Science Foundation Grants PHY-0757058 and PHY-0823459. This material is based upon work supported by NSF’s LIGO Laboratory which is a major facility fully funded by the National Science Foundation. Alongside previously mentioned software packages we would also like to thank the following codes for their use in this study: \texttt{SCIPY} \cite{virtanen2020scipy}, \texttt{NUMPY} \cite{harris2020array}, \texttt{PANDAS} ~\cite{Hunter2007, mckinney-proc-scipy-2010}, \texttt{MATPLOTLIB} \cite{reback2020pandas}, and \texttt{GWPY} \cite{gwpy}. This work has been allocated LIGO document number P2200229, and Virgo document number VIR-0777A-22.
PR is supported by STFC grant ST/S505328/1. VR is supported by STFC grants ST/V001396/1 and ST/V00154X/1. IH is supported by STFC grants ST/T000333/1 and ST/V005715/1. MD acknowledges the support from the Amaldi Research Center funded by the MIUR program 'Dipartimento di Eccellenza' (CUP:B81I18001170001) and the Sapienza School for Advanced Studies (SSAS). ST is supported by Swiss National Science Foundation (SNSF) Ambizione Grant Number: PZ00P2$\_$202204. AM acknowledges the support of the European Gravitational Observatory under the convention EGO-DIR-63-2018

\section{Appendix - Supplementary figure}\label{appendix1}

\begin{figure}[t!]
    \centering
    \includegraphics[width=\linewidth]{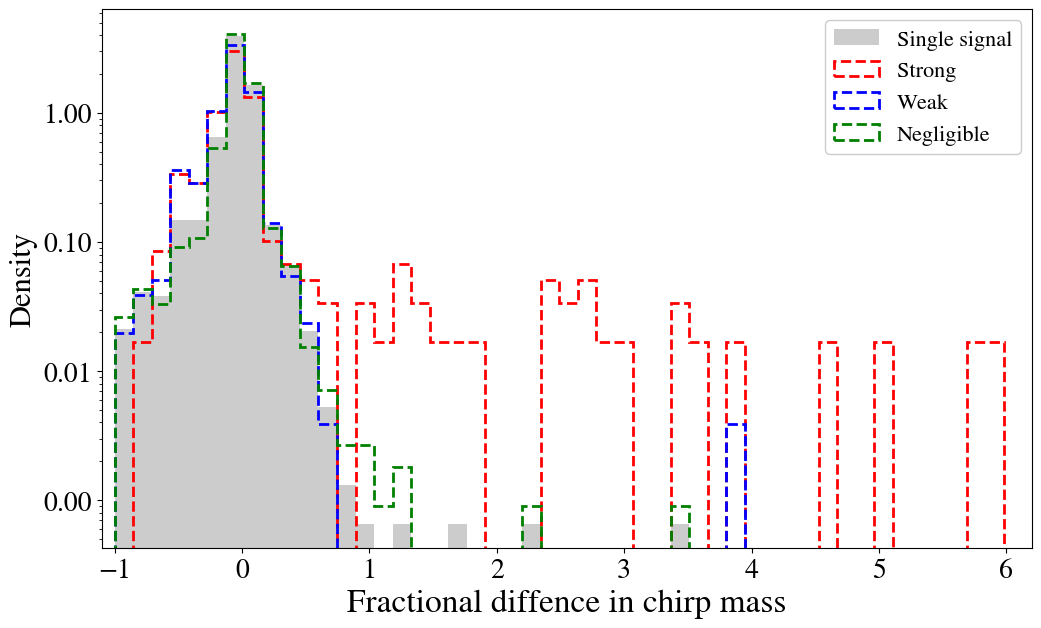}
    \caption{Distribution, across the overlap regions, of the fractional difference between the recovered and injected chirp masses for \texttt{PyCBC}) triggers in the PAIRS injection set. Also shown, in grey, is the same distribution for recovered and injected in the SINGLES injection set. A perfect pipeline, with an infinitely finely gridded template bank, would find injections with a delta spike on zero in this plot. Here the distribution is shown as a function of the density of signals, rather than the signal numbers.}
    \label{fig:delta_Mc_density}
\end{figure}

The plot shown in Figure \ref{fig:delta_Mc_density} is similar to that in Figure \ref{fig:delta_Mc}, however, the y axis has been adjusted to show the normalised distributions, rather than the number of triggers. This allows more direct comparison between the four distributions. 

\bibliographystyle{unsrt}
\bibliography{main.bib}
\end{document}